# Effect of CLIQ on training of HL-LHC quadrupole magnets

S. Stoynev, G. Ambrosio, K. Amm, J. DiMarco, S. Feher, P. Ferracin, V. Marinozzi, S. Prestemon, A. B. Yahia

*Abstract—* The high-luminosity LHC upgrade requires stronger than LHC low-beta quadrupole magnets to reach the luminosity goals of the project. The project is well advanced and HL-LHC quadrupole magnets are currently being commissioned in US Labs (MQXFA magnets) and CERN (MQXFB magnets). Those are the first Nb$_3$Sn magnets to be used in any large particle accelerator. At development stages, many Nb$_3$Sn accelerator sub-scale models showed relatively slow training and MQXFA magnets were projected to have low tens of quenches before reaching operational field. Recently it was shown that dedicated capacitor-based devices can affect Nb$_3$Sn magnet training, and it was suggested that CLIQ, a capacitor-based device intended for quench protection, can do too. The present paper investigates effects on training likely induced by CLIQ, using the base fact that only half the coils in a quadrupole experience upward current modulation at quench because of capacitor discharge. The study encompasses all MQXFA production magnets trained at BNL to date. No other high-statistics data from identical magnets (series) with CLIQ protection exist so far. Implications and opportunities stemming from data analysis are discussed and conclusions drawn.

## I. INTRODUCTION

STATISTICAL analyses were successfully applied in past to assess performance and behavior of superconducting accelerator magnets [1]-[3]. In most cases those rely on relatively large data samples whereas cutting-edge developments naturally have limited statistics to start with. At some point data accumulation is enough to start making meaningful observations. We are entering this stage with Nb$_3$Sn quadrupole magnets being commissioned as part of the HL-LHC upgrade program [4]. In the USA, the Accelerator Upgrade Project (AUP) [5] is responsible for delivering 4.2-m-long MQXFA quadrupole magnets (and cryo-assemblies), whereas in Europe longer but otherwise similar MQXFB magnets are fabricated and tested. Multiple MQXFA magnets were already tested and, as each magnet contains four coils, this is becoming a sizable data sample. One of the main goals of commissioning is to "train" the Nb$_3$Sn magnets to predefined current level suitable for operations in HL-LHC. It was expected that MQXFA training could be long [6]. As part of the quench protection system all those magnets use CLIQ [7] which is a capacitor-based device modulating the current through the magnet at quench detection. By design, part (half) of the coils in a magnet see upward modulation (at quench) and the rest – downward modulation. Thus, the current soon after quenching is temporarily much higher in half the coils. Recently another capacitor-based device, QCD, was developed and tested [8] with the explicit goal to affect magnet training. It was shown that the test object in multiple thermal cycles, a Nb$_3$Sn short dipole coil in a "mirror" magnet configuration, repeatedly did not train when QCD was used and did train when QCD was not used. Quoting [9], where the case for "coil training" and not "magnet training" was made, the QCD paper suggested that just as QCD, which is designed to boost the current in all coils in a magnet, can affect training in coils, CLIQ can do so as well.

The present paper uses all MQXFA training test data, taken at the Brookhaven National Laboratory (BNL), to-date to perform statistical analysis and further studies on them. It aims to find out if coils in different quadrants of the magnet train the same and investigates statistically significant differences observed. We suggest explanations for results obtained and, following the analysis, propose modifications to superconducting magnet/sample testing, for low temperature superconducting magnets.

## II. MQXFA MAGNETS

### A. Magnets and Testing Conditions

After cable and coil fabrication and magnet assembly, jointly performed by BNL, Lawrence Berkeley National Laboratory (LBNL) and Fermi National Accelerator Laboratory (FNAL) teams, MQXFA magnets are delivered and tested at BNL in their vertical magnet test facility. Specific test plans are agreed on within the AUP and the HL-LHC management at CERN, with the goal to test the magnets for meeting pre-defined objective markers. One of those is training the superconducting magnets to certain "acceptance" current, $I_a$ = 16530 A, which is 300 A above the "nominal" current $I_n$ = 16230 A. MQXFA magnets with numbers 03, 04, 05, 06, 07, 08, 10, 11, 13 were pre-stressed to levels able to support the coils (no pole-coil separation) to $I_n$ and tested. All of them had full training sequences at 1.9 K except magnet MQXFA06 which started

Submitted for review September 21, 2023

This work was supported by the U.S. Department of Energy, Office of Science, Office of High Energy Physics, through the US LHC Accelerator Upgrade Project (AUP). *(Corresponding author: Stoyan Stoynev, stoyan@fnal.gov).*

S. Stoynev, G. Ambrosio, J. DiMarco, S. Feher and V. Marinozzi are with the Fermilab National Accelerator Laboratory, Batavia, IL 60510 USA.

K. Amm and A. B. Yahia are with Brookhaven National Laboratory, Upton, NY 11973 USA.

P. Ferracin and S. Prestemon are with Lawrence Berkeley National Laboratory, Berkeley, CA 94720-8203 USA.

Color versions of one or more of the figures in this article are available online at http://ieeexplore.ieee.org

training at 4.5 K before changing to 1.9 K [10]. The present study is mainly concentrated on data up to nominal current, but we add 10 A to define our analysis threshold current $I_{th} = I_n + 10$ A = 16240 A. On one hand the magnets were pre-stress "tuned" for "nominal" performance and on the other hand all magnets reached nominal current + 10 A [10]-[12]. Two coils in MQXFA05 ("Q1" and "Q3") were the only ones which quenched within 10 A above $I_n$, and for Q3 it was the first quench. MQXFA07 exceeded $I_{th}$ only at 100 A/s ramp rate, 20 A/s being the nominal ramp rate for all training ramps and was limited by a "Q3" coil for all 1.9 K quenches. This Q3 coil was found to be "weak", as defined later, and was not included in the bulk of the analysis.

The four coils in a magnet are placed in pre-defined quadrants, Q1 to Q4 [13], with transport current flowing in the direction Q1-Q4-Q2-Q3. CLIQ [7], which is part of the magnet protection system, has its two leads connected at the Q1-Q4 (positive) and Q2-Q3 (negative) junctions. Thus, when CLIQ is discharged its current is divided between two sets of coils: coils in Q2/Q4 experience "overcurrent" – coil current higher than at the moment of discharge, Fig.1 - due to addition of power supply and CLIQ currents; coils in Q1/Q3 experience "undercurrent" due to currents flowing in opposite directions. CLIQ is used in all training quenches and high-current trips with unchanged parameters (voltage, capacitance).

The choice of placing a coil in a given quadrant is based on coil RRR information, peak voltage development in normal and abnormal conditions, coil sizing observations, availability of FNAL/BNL-fabricated coils with a soft preference to have two of each in a magnet.

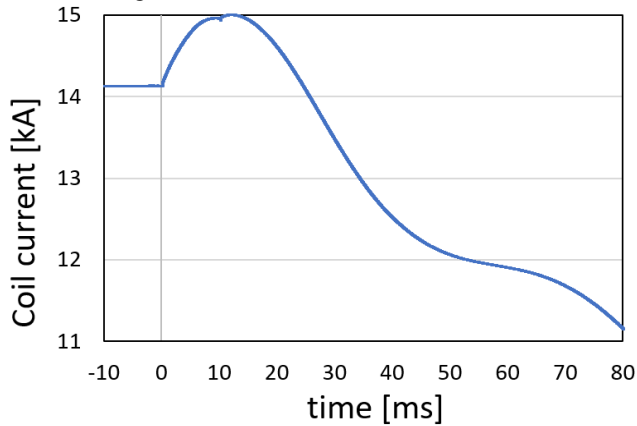

**Fig. 1.** Current through coils with "overcurrent" – data from the first ramp (quench) in MQXFA13. Quench is detected at "0 ms" and CLIQ is discharged without delay causing the current to peak in ~ 12.5 ms. At the 10 ms mark the dump resistor switch is opened which is seen as a small disturbance on the plot.

*B. Training Data*

The data of interest is "coil" training, as noted in [9] – we count and analyze training quenches in each coil independently on others. We only look at training sequences where coils were never tested before. One subsequent magnet test, MQXFA08b, contained a new coil, a substitute, and the training of this coil is included in the analysis. Most of the quench current "gains", i.e., gains in current reached per ramp/quench, are small, averaging to about 100-200 A. However, we have reasons to suspect [8] that levels of current gains may matter for this analysis. Given CLIQ upward peak-current is ~ 900 A, Fig. 1, and that 300 A is the "safe" margin we use in training (i.e. $I_a - I_n$) we study three inclusive tiers of data: number of quenches with gain < 300 A; with gain < 2 x 300 A; with no constraints (in practice the same as gain < 3 x 300 A). In all cases, the very first quench, or high current trip (like in MQXFA11) in a magnet is nominally excluded, as CLIQ only acts at high current after it; also, counting the gain is with respect to the maximum power supply current provided previously to the coil. Counting of quenches is carried to $I_{th}$. Further, we consider the existence of "weak" coils which we define as coils that experienced quench-to-quench detraining of > 300 A (i.e., they lose at least 300 A in a consequent training quench) eventually reaching $I_{th}$ later. Ideally, coils train with no detraining quenches and large detraining does not give a good base for training studies and comparisons. Finally, coils were fabricated either at FNAL (numbers start with 1) or BNL (numbers start with 2). Data, as described, are presented in Table I.

We find that "non-weak" coils are 34 out or 37 ("all"). Thus, our parameter space and updated data tiers to analyze are **(coil condition, quench current gain condition)** with two and three options for the two dimensions, respectively.

TABLE I
MQXFA MAGNET QUENCH DATA

| Magnet | Q1 | Q2 | Q3 | Q4 |
|---|---|---|---|---|
| MQXFA03 | 204<br>0/0/0<br>0 | 110<br>0/0/0<br>0 | 202<br>0/0/0<br>0 | 111<br>4/7/8<br>1 |
| MQXFA04 | 203<br>1/1/1<br>0 | 113<br>3/4/4<br>1 | 206<br>0/0/0<br>0 | 112<br>0/0/0<br>0 |
| MQXFA05 | 207<br>3/4/4<br>0 | 116<br>0/0/1<br>0 | 209<br>1/1/1<br>0 | 115<br>0/0/0<br>0 |
| MQXFA06 | 122<br>1/1/1<br>0 | 119<br>0/0/0<br>0 | 211<br>0/2/2<br>0 | 123<br>0/0/0<br>0 |
| MQXFA07 | 212<br>1/1/1<br>0 | 124<br>0/0/0<br>0 | 214<br>6/7/7<br>1 | 114<br>0/0/0<br>0 |
| MQXFA08 | 215<br>0/0/0<br>0 | 126<br>0/0/1<br>0 | 213<br>0/0/0<br>0 | 128<br>0/0/0<br>0 |
| MQXFA08b | - | - | 219<br>2/2/3<br>0 | - |
| MQXFA10 | 132<br>8/8/8<br>0 | 221<br>0/0/0<br>0 | 131<br>5/6/6<br>0 | 129<br>0/0/0<br>0 |
| MQXFA11 | 223<br>1/1/1<br>0 | 222<br>0/0/0<br>0 | 134<br>5/6/6<br>0 | 135<br>0/0/0<br>0 |
| MQXFA13 | 227<br>14/14/14<br>0 | 139<br>0/0/1<br>0 | 229<br>3/4/4<br>0 | 141<br>0/0/0<br>0 |

For each magnet and quadrant there are three rows; the top one shows the coil number in the quadrant; the second one shows number of quenches to threshold current at 1.9 K with the three tiers of data ordered as explained in the text; the bottom row shows number of detraining quenches with current loss above 300 A - "weak" coils with such non-zero detraining quenches are highlighted. For MQXFA08b, only the new coil, a substitute, is shown.

For MQXFA06, where training started at 4.5 K [10], we exclude all 4.5 K training quenches before the 1.9 K tests. Training continued at 1.9 K. Comparing training at different temperatures requires accounting for conductor current limits and proximity to critical surface at least, it unnecessary complicates analysis. It is worth mentioning that the measured Short Sample Limits at 1.9 K for the conductor on a coil-basis were between 21.2 kA and 22.8 with a mean of 22.2 kA and a standard deviation of 0.2 kA. RRR measurements also yielded relatively small standard deviations among coils. No dependencies between those and training performance were found so far. No dependencies between training performance and other fabrication variables, like choice of positioning a coil in a given quadrant, were found.

For the purposes of our analysis, we only consider if a coil quenches at all, for each data tier, not the exact number of quenches which are shown in Table I for completeness.

There were no quenches in Q1/Q3 with gains above 600 A. There were three quenches in Q2/Q4 with gain above 600 A, all of them as the second magnet quench: +891 A (Q2 in MQXFA05), +789 A (Q2 in MQXFA08) and +663 A (Q2 in MQXFA13). Quench current "gain" with respect to the maximum CLIQ-induced peak current in the preceding quench, for the three cases, was -1 A, -92 A and -211 A, respectively. Thus, the quench currents were lower than the preceding peak currents. The overcurrent in quench #1 of magnet 13 was above the quench #2 current (14789 A) for about 12.5 ms, Fig. 1.

There were no other training quenches to $I_{th}$ in Q2/Q4 at 1.9 K, excluding (two) "weak" coils and "first" quenches in a magnet.

In MQXFA10 there were two quenches in Q2 above $I_{th}$ and below $I_a$, with gains of + 55 A and – 5 A (loss), respectively. We address this exceptional case separately later, there were no other Q2/Q4 quenches in that current range for non-weak coils.

The following coils quenched first in a magnet: Q2 in MQXFA03/04/05/06/08 and Q4 in MQXFA07; Q1 in MQXFA13 and Q3 in MQXFA10. No quenches (to threshold current) occurred in Q2/Q4 after they first quenched in a magnet and there were multiple quenches in Q1/Q3 after they first quenched in a magnet. A high current trip, with usual quench protection settings, occurred in MQXFA11 before any training quench, the following quench is counted in the analysis.

Coils in Q3 and Q1, in MQXFA08 and MQXFA13, respectively, gradually detrained after reaching $I_{th}$; according to the rules we set they are included in the analysis.

### III. Data Analysis

*A. Setting up questions and framework to address them*

As we have indications from elsewhere [8], as well as accumulating suspicions from observations, the immediate question is this: <u>*do coils in Q1/Q3 train differently than coils in Q2/Q4?*</u> Counting of quenched coils fulfills conditions for Bernoulli trials except the probability of coil *i* (out of *n*) to quench at least once at given conditions, $p_i$, may be coil-dependent. Its sample average is $\bar{p} = \frac{\sum p_i}{n}$, variance is $\sigma_p^2 = \frac{1}{n}\sum(p_i - \bar{p})^2$; all sums run to *n*. Resolving the latter equation leads to $\sum p_i^2 = n(\sigma_p^2 + \bar{p}^2)$. Then the variance of the distribution from independent Bernoulli trials can be written as $Var = \sum p_i(1-p_i) = n\frac{\sum p_i}{n} - \sum p_i^2 = n\bar{p}(1-\bar{p}) - n\sigma_p^2$.

Thus, *Var* is not larger than the variance of the Binomial distribution constructed from the average probability. Then the Binomial distribution is still a good base for setting limits in analyses even if the quench probability changes coil-to-coil.

The kind of questions we are asking is addressed by *hypothesis testing* and we benefit from well-developed apparatus elsewhere [14]-[17]. In our case we deal with 2x2 contingency tables and are looking for possible relations between categories. For instance, counting quenched, at least once, "non-weak" coils with gain < 600 A gives us fractions of coils quenched, f(Q1||Q3) = 0.72 and f(Q2||Q4) = 0.00. The corresponding 2x2 contingency table extracted from Table I is:

|  |  | Q1∥Q3 | Q2∥Q4 |
|---|---|---|---|
| unquenched |  | 5 | 16 |
| quenched |  | 13 | 0 |

and we define the notation **5/13 vs 16/0** for it.

Our Null Hypothesis, H0, is that there is no difference between the two categories (quadrant sets) of coils, i.e., there is no relation between quenching and coil placement in a given pair of quadrants. We want to test this hypothesis, to start with.

For 2x2 contingency tables there are two "exact" statistical tests which are particularly relevant for small statistical samples – Fisher's exact test, FT, and Barnard's exact test, BT (see [15]). The "exact" tests have different boundary conditions and yield the "exact" results satisfying them. The result is the p-value (PV), a partially integrated probability distribution, giving the probability for the test to wrongly reject H0. Fisher's test is often considered to be more conservative as its boundary conditions are more constraining.

The R Project for Statistical Computing [18] is utilized for calculations of p-values and Confidence Intervals (CI). In particular, we use the functions fisher.test(x) and BarnardTest(x, method = "z-pooled") from DescTools[19]. Those were also assessed with literature test examples.

We give results for p-values as "one-sided" (1-s) and "two-sided" (2-s) for completeness. For most tests we argue 1-s PVs are more appropriate - when we have a reason to believe the Null Hypothesis may be wrong in a particular way.

Depending on the field of research and certainty needed, a rejection threshold on the p-value, a.k.a. significance level (SL), is chosen very high (5% is often quoted) or very low (e.g., 2 x 3 x $10^{-7}$ a.k.a. "5σ" requirement) [20]. We set SL = 0.1% with the caveat it is for purposes of discussion. We utilize the "odds ratio" (OR) - if the ratio of quenched and unquenched coils per category is the "odds" for that category, then the ratio of "odds" between the two categories in a 2x2 table is the OR. The Null Hypothesis is equivalent to OR ≡ 1. CIs show the likely range where the "true" OR is with declared confidence/certainty based on observed data and test statistic. CI gives complimentary to PV information: distance of CI (range) from OR = 1 informs us about the magnitude of the "rejection";

narrower CI implies higher statistical power (probability to reject H0 if false) and more reliable estimate.

*B. Quenching in Q1/Q3 vs Q2/Q4*

From Table I one can find conditions for Q1||Q3 vs Q2||Q4 quench data and perform tests outlined assessing H0: there **is no** relation between quenching and where coils are placed in a magnet (Q1||Q3 vs Q2||Q4). Results are presented in Table II. Subjectively, we expect Q2||Q4 to quench less and the 1-s PV is more appropriate than the 2-s PV. Small PVs across the board in Table II indicate we have reasons to doubt H0, and our random choice of SL = 0.1% is enough to **reject** it for all but the two data tiers with "no limit" quench current gain. Including "weak" coils does not change the perspective much because they are a small fraction of all coils (<10%). The table suggests that current gain magnitude affects results through a threshold. There is not much difference in results with 300 and 600 A gain thresholds because we eventually count all training quenches – remember that most gains are in the range 100-200 A; if we set a too low threshold we do not count some of the quenches. But behavior abruptly changes for the case above 600 A making it special. The data tier which produces the lowest PV is **("non-weak", "< 600 A")**, and the PV is much lower than our adopted SL (i.e., the rejection threshold). For that tier the 95% CI is narrow and away from OR = 1, based on FT, Table II. The Alternative Hypothesis, which is to be accepted if H0 is rejected for a given tier, is H1: there **is** relation between quenching and where coils are placed in a magnet (Q1||Q3 vs Q2||Q4).

To assess the importance of excluding the "first" quench from the analysis, we re-count quenching coils in the **("non-weak", "< 600 A")** tier explicitly **counting "first" quenches in a magnet if they occurred in those coils**. Results are in the last row of Table II and show that including the first magnet quenches in the analysis increases PVs by orders of magnitude!

Magnet MQXFA14b was also trained [11] but with increased prestress levels and is not included in the analysis for simplicity.

TABLE II
HYPOTHESIS TESTING FOR QUENCHING OF COILS IN Q1||Q3 VS Q2||Q4

| Conditions and Data | Test | PV (1-s) | PV(2-s) | 95% CI |
|---|---|---|---|---|
| **("non-weak", "<300 A"); 6/12 vs 16/0** | FT<br>BT | 3.4 x 10$^{-5}$<br>2.1 x 10$^{-5}$ | 3.7 x 10$^{-5}$<br>4.1 x 10$^{-5}$ | (0.0,0.15) |
| **("non-weak", "<600 A"); 5/13 vs 16/0** | FT<br>BT | 9.2 x 10$^{-6}$<br>3.6 x 10$^{-6}$ | 9.8 x 10$^{-6}$<br>7.2 x 10$^{-6}$ | (0.0,0.12) |
| ("non-weak", "no limit"); 5/13 vs 13/3 | FT<br>BT | 2.4 x 10$^{-3}$<br>9.9 x 10$^{-4}$ | 2.6 x 10$^{-3}$<br>1.9 x 10$^{-3}$ | (0.0,0.44) |
| ("all", "<300 A"); 6/13 vs 16/2 | FT<br>BT | 4.7 x 10$^{-4}$<br>1.7 x 10$^{-4}$ | 6.3 x 10$^{-4}$<br>2.8 x 10$^{-4}$ | (0.0,0.32) |
| ("all", "<600 A"); 5/14 vs 16/2 | FT<br>BT | 1.4 x 10$^{-4}$<br>4.8 x 10$^{-5}$ | 1.9 x 10$^{-4}$<br>8.2 x 10$^{-5}$ | (0.0,0.26) |
| ("all", "no limit"); 5/14 vs 13/5 | FT<br>BT | 6.4 x 10$^{-3}$<br>3.9 x 10$^{-3}$ | 8.6 x 10$^{-3}$<br>7.7 x 10$^{-3}$ | (0.0,0.58) |
| **("non-weak", "<600 A" + first quenches); 5/13 vs 10/6** | FT<br>BT | 4.50 x 10$^{-2}$<br>3.09 x 10$^{-2}$ | 8.6 x 10$^{-3}$<br>7.7 x 10$^{-3}$ | (0.0,0.97) |

For each data condition, we show the 2x2 contingency table in the notations introduced earlier in the text. FT and BT (tests) are performed giving 1-s and 2-s PV and 95% CI (FT, 1-s based). Non-rejected 1-s H0's are highlighted.

It quenched first in Q2 (15857 A) and then three times in Q3 above $I_{th}$ before reaching $I_a$. It is in line with the observations in other magnets - no Q2/Q4 training quenches to operational levels after the first magnet quench.

*C. Conditional/subset Hypothesis Testing*

Dividing data in sub-categories gives us more insights but statistical power is reduced. We investigate four more H0's, all for the tier **("non-weak", "< 600 A")** : (I) a first quenched coil in a magnet has other training quenches independently on where it is in the magnet (Q1||Q3 vs Q2||Q4); (II) for quenches in Q1||Q3 only: there is no relation between quenching and coil origin (FNAL coils vs BNL coils); (III) for BNL coils only: there is no relation between quenching and where coils are placed in a magnet (Q1||Q3 vs Q2||Q4); (IV) for FNAL coils only: there is no relation between quenching and where coils are placed in a magnet (Q1||Q3 vs Q2||Q4). Results are in Table III.

While testing of (I), (II) and (III) could have given valuable information they actually demonstrate how poor statistical power makes them almost irrelevant – for (II) and (III) the 95% CIs are so wide that they contain the whole meaningful OR range: [0,1].

Testing of (IV) however yields a quite important result – this hypothesis is to be rejected and thus the differences in quenching between quadrants are not explained by differences in coil origin!

TABLE III
HYPOTHESIS TESTING (SUBSETS)

| Hypothesis and data | Test | PV (1-s) | PV(2-s) | 95% CI |
|---|---|---|---|---|
| **(I) ; 0/2 vs 6/0** | FT<br>BT | 3.6 x 10$^{-2}$<br>1.1 x 10$^{-2}$ | 3.6 x 10$^{-2}$<br>1.1 x 10$^{-2}$ | (0.0,0.79) |
| **(II) ; 0/4 vs 5/9** | FT<br>BT | 0.23<br>0.11 | 0.28<br>0.20 | (0.0,2.83) |
| **(III) ; 5/9 vs 2/0** | FT<br>BT | 0.18<br>0.12 | 0.18<br>0.12 | (0.0,2.56) |
| **(IV) ; 0/4 vs 14/0** | FT<br>BT | 3.3 x 10$^{-4}$<br>7.2 x 10$^{-5}$ | 3.3 x 10$^{-4}$<br>7.2 x 10$^{-5}$ | (0.0,0.13) |

For each hypothesis, we show the 2x2 contingency table in the notations introduced earlier in the text. Testing and results follow notations in Table II.

*D. "Weak" Coils and Future Performance*

According to our initial expectations there had to be no training quenches in Q2||Q4 below some current gain threshold. Formally this is what we observe counting to $I_{th}$. Nevertheless, we want to know why there are two seemingly training quenches in Q2 of MQXFA10 above that current level – at 16253 A and 16525 A. The latter is a -5 A detraining quench with respect to the previously reached $I_a$. Both quench currents are high with respect to operational conditions and one can argue lack of enough mechanical support there could lead to quench current instabilities. However, more direct argument is the frequency of detraining quenches observed. Several coils were already declared "weak" because of large detraining (> 300 A). In fact, every magnet, except MQXFA05, saw at least one coil with detraining quenches. Even in MQXFA10 there was another coil, beyond Q2, temporarily detraining [10]. Earlier QXF prototypes showed similar detraining behavior [4], so detraining quenches in MQXF coils are common. It is

plausible that both quenches we saw in Q2 of MQXFA10 were detraining quenches – detraining with respect to the coil conductor limit or the "effective" current [8] reached after CLIQ boosting, whatever is lower. The coil with its two detraining quenches is not exceptional with respect to other (operable) coils in the series. However, large detraining with respect to the "effective" current may forebode recurring weakness at operating conditions. Indeed, both MQXFA03 and 04 quenched once (though only once) in their "weak" coils in horizontal testing of the cryo-assembly they were put in [21].

*E. Discussion*

The data analysis suggests that Q2‖Q4, unlike Q1‖Q3, coils do not train, and this is not related to their fabrication origin. Data suggest there is a quench current gain threshold, 600 A or more, that starts to affect the absolute validity of those observations. Adding the first quench in a magnet to the analysis drastically changes the statistical outcome.

The quadrant differentiation, the quench current gain threshold, and the explicit importance of the first quench in a magnet in the analysis are all features relatable to CLIQ and its effects on the magnets. This is the expected behavior if we assume the training is affected by CLIQ according to [8]. The only unknown is the exact quench current gain threshold where the boost-effect breaks ("effective current boost"). As described, the lowest gain among the quenches with gain > 600 A followed current boost from the previous quench shown on Fig. 1; the quench occurred 211 A below the peak current seen on the figure. One of the other two high gain quenches occurred at the level of the boost current within 1 A. If there is no "detraining" involved, and conductor limit is higher than the previously reached peak, the 211 A "dip" must be related to the "effective current boost", which is potentially different than the "peak current boost". The effective quantity depends on the time for which there is "overcurrent". A quench will not occur below it without detraining but could develop well above it, as for any other training quench. Currently, due to many variables and unknowns involved, an answer to what "effective" is can only be given statistically. We do not have the statistical power to resolve it for MQXFA coils.

If we accept that CLIQ affects coil training as described, then we also support the statement made in [9] that, at least in first approximation, coils in magnets can train independently. In this context, "magnet training" is a misleading term.

## IV. MAGNET TRAINING OPTIMIZATION WITH CLIQ

A magnet/s and its quench protection system, particularly CLIQ, could be optimized in terms of quench performance. The most straightforward way is to alter CLIQ polarity after each training quench or in a succession of stair-steps-like high-current trips, a.k.a. "quenchless" training [8]. We investigate feasibility of swapping CLIQ polarity in MQXFA magnets.

Fig. 2 (left) shows the maximum coil-to-ground voltage for each coil-turn in a magnet as obtained by STEAM-LEDET simulation [22], [23], MQXFA04 coil parameters were used at nominal current. The cases with "nominal" and "reversed" CLIQ polarity are compared. As seen, reversing CLIQ does not worsen overall conditions. We also observe that maximum voltages turn-to-turn do not change significantly by reversing the CLIQ polarity. However, protection heater and CLIQ heater relative voltages depend on their relative polarities. The magnet protection is designed such that those voltages are minimized, and Fig 2 (right) shows the worst case of heater-to-coil voltage development over temperature. A small initial peak, related to CLIQ and heater discharging, is followed by voltage development driven by coil resistance increase due to Joule heating mostly. This second part is largely independent on CLIQ polarity and is the critical one for insulation strength as seen by break-down curves for the relevant heater-to-coil distance. If the CLIQ polarity is reversed the fast decaying (and oscillating) CLIQ voltage in effect is added to the heater voltage instead of being subtracted. The effect will vary significantly turn-to-turn but will dominate lower temperatures before CLIQ power is fully exhausted and corresponding heat dissipated. While we have not completed those studies, which are more complex than the coil-to-ground considerations, initial indications are reassuring. We assess that CLIQ polarity flip will ultimately have no detrimental effect on magnet safety and is a viable option during testing. We acknowledge the need of more detailed studies which we intend to conduct in future.

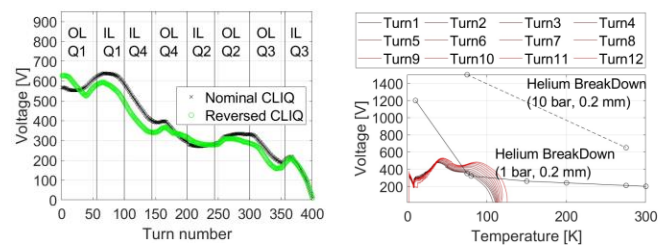

**Fig. 2.** Simulation results. Left: maximum voltages to ground in coil-turns (outer layers, OL; inner layers, IL) for "nominal" and "reversed" CLIQ polarity. Right: coil-turn-to-heater voltage development (worst layer) and breakdown voltage contours in He gas for relevant distance, "nominal" case.

## V. CONCLUSION

Following statistical analysis of all trained MQXFA magnets to date, we found that coils which were subjected to "overcurrent" due to CLIQ did not train. The analysis showed there were few exceptions related to "weak" coils or detraining quenches, albeit at above nominal current for the latter. Further analysis on the role of quench current gain thresholds, magnet first quenches and fabrication origin of coils revealed that with high likelihood the lack of training in particular coils is caused by CLIQ. No other plausible and probable explanation for all observations is available at present.

We investigated and demonstrated by simulations that reversing CLIQ polarity after each quench during training, or in a succession of high-current stair-step trips, is potentially a safe training reduction method to apply to MQXFA magnets. Designing magnet protection to accommodate changing CLIQ polarities in general can reduce magnet/coil training time substantially, depending on the magnitude and duration of the applied current boost.